\documentclass[10pt]{article}
\usepackage{graphicx}
\usepackage{amsmath}

\def\Title#1{\begin{center} {\Large #1 } \end{center}}
\def\Author#1{\begin{center}{ \sc #1} \end{center}}
\def\Address#1{\begin{center}{ \it #1} \end{center}}

\newcommand\pubblock{\rightline{\begin{tabular}{l} Proceedings of the Second Annual LHCP\\ \pubnumber\\
         \pubdate  \end{tabular}}}

\newenvironment{Abstract}{\begin{quotation} \begin{center} 
             \large ABSTRACT \end{center}\bigskip 
      \begin{center}\begin{large}}{\end{large}\end{center} \end{quotation}}

\newenvironment{Presented}{\begin{quotation} \begin{center} 
             PRESENTED AT\end{center}\bigskip 
      \begin{center}\begin{large}}{\end{large}\end{center} \end{quotation}}

\def\beq{\begin{equation}}
\def\eeq#1{\label{#1}\end{equation}}
\def\eeqn{\end{equation}}

\def\beqa{\begin{eqnarray}}
\def\eeqa#1{\label{#1}\end{eqnarray}}
\def\eeqan{\end{eqnarray}}

\let\bar=\overbar

\def\Dslash{\not{\hbox{\kern-4pt $D$}}}
\def\dslash{\not{\hbox{\kern-2pt $\del$}}}

\def\msb{{\bar{\ssstyle M \kern -1pt S}}}

\usepackage{color}

\newcommand{\veslt}{\ensuremath{{\hbox{$\vec{E}$\kern-0.60em\lower-.05ex\hbox{/}\kern0.15em}}_{T}}}   
\newcommand{\eslt}{\ensuremath{{\hbox{$E$\kern-0.60em\lower-.05ex\hbox{/}\kern0.15em}}_{T}}}   
\newcommand{\zmumu}{\ensuremath{\mathrm{Z}\rightarrow\mu\mu}}
\newcommand{\wenu}{\ensuremath{\mathrm{W}\rightarrow \mathrm{e}\nu}}

\newcommand{\sig}{\ensuremath{\mathcal{S}}}
\newcommand{\pchisq}{\ensuremath{\mathcal{P}_2(\sig)}}
\newcommand{\vet}{\ensuremath{\vec\varepsilon}}
\newcommand{\like}{\ensuremath{{\cal L}}}
\newcommand{\signif}{\ensuremath{{\mathcal S}}}

\newcommand\pubnumber{ CMS CR-2014/179 }

\newcommand\pubdate{\today}

\def\affiliation{
On behalf of the CMS Experiment, \\
Department of Physics \\
Cornell University, Ithaca, NY 14853, U.S.A }

\begin{document}

\large
\begin{titlepage}
\pubblock

\vfill
\Title{ Missing transverse energy significance at CMS }
\vfill

\Author{ Nathan Mirman, Yimin Wang, and James Alexander }
\Address{\affiliation}
\vfill
\begin{Abstract}

Missing transverse energy significance may be used to help distinguish real missing transverse energy due to undetected particles from spurious missing transverse energy due to resolution smearing. We present a description of the missing transverse energy significance variable, and assess its performance in \zmumu, dijet, and \wenu\ events using the CMS 8 TeV dataset.

\end{Abstract}
\vfill

\begin{Presented}
The Second Annual Conference\\
 on Large Hadron Collider Physics \\
Columbia University, New York, U.S.A \\ 
June 2-7, 2014
\end{Presented}
\vfill
\end{titlepage}
\def\thefootnote{\fnsymbol{footnote}}
\setcounter{footnote}{0}
%

\normalsize 


\section{Introduction}

In many analyses at the LHC, it is important to distinguish between events with a genuine source of missing transverse energy, denoted \veslt, arising from undetected particles, and those with spurious \veslt\ due to object misreconstruction, finite detector resolution, or detector noise.  We have developed the \veslt\ significance variable, \signif, to help make this distinction on an event-by-event basis.  It is based on \veslt\ reconstructed with a particle flow method at CMS \cite{Chatrchyan:2011tn}.  Here we give a description of \veslt\ significance and evaluate its performance in \zmumu, dijet, and \wenu\ events using the full CMS 8 TeV dataset.

The \veslt\ significance is defined as the log-likelihood ratio:
\begin{equation}
  \signif \equiv 2\ln\left(\frac
  {\like(\vet=\sum\vet_{i})}
  {\like(\vet=0)}
  \right),
  \label{eq:metsig:defn}
\end{equation}
where \vet\ is the true \veslt, and $\sum\vet_i$ is the observed \veslt, computed by summing over all reconstructed objects in the event.  In the numerator, we evaluate the likelihood that the true value of \veslt\ equals the observed value, while the denominator corresponds to the null hypothesis (that the true \veslt\ is zero).

When the likelihood $\like(\vet)$ is a Gaussian, the significance can be expressed in a more compact form:
\begin{equation}
  \signif = \Big(\sum\vet_i \Big)\!^\dag {\mathbf V^{-1}} \Big( \sum\vet_i \Big),
  \label{e:metsig-gaussian}
\end{equation}
where ${\mathbf V}$ is a $2\times 2$ covariance matrix.  Here, \signif\ is simply a $\chi^2$ variable with two degrees of freedom.

The total \veslt\ resolution captured in covariance $\mathbf{V}$ is primarily determined by the hadronic activity in each event.  For \veslt\ significance, this includes jets with $p_T > 20$ GeV, and objects with $p_T < 20$ GeV that make up the unclustered energy of the event.  All jets passing the 20 GeV threshold enter into the total covariance with a resolution of the form
\begin{equation}
  \mathbf{U} = \left( \begin{matrix}
    \sigma_{p_{T}}^2 & 0 \\
                 0 & p_{T}^2\,\sigma_{\phi}^2
  \end{matrix} \right).
  \label{eq:cov}
\end{equation}
The quantities $\sigma_{p_{T}}$ and $\sigma_{\phi}$ are determined in simulation and are then retuned with data using five $\eta$-dependent scale factors, so that $\sigma(p_T,\eta) = a(\eta)\times\sigma^{\text{sim}}$.  The values of $a(\eta)$ are determined in a likelihood fit in the \zmumu\ channel, where we maximize the null hypothesis (that each event has zero true \veslt).

All objects falling below the 20 GeV threshold are contained in the unclustered energy of the event, where the momentum of the unclustered energy is a vectorial sum over all of its constituents.  The resolution of the unclustered energy is parameterized by the scalar $p_T$ sum of its constituents:
\begin{eqnarray}
  \sigma_{uc}^2 &=& \sigma_0^2 + \sigma_s^2\sum_{i=1}^{n}{|\vec{p}_{T_i}|},
  \label{e:pseudo}
\end{eqnarray}
where the values of $\sigma_0^2$ and $\sigma_s^2$ are determined in the \zmumu\ channel likelihood fit described above.  The resolution of the unclustered energy is assumed to be isotropic in the transverse plane of the detector.

Electrons and muons are assumed to have negligible resolutions when compared to the hadronic component of each event, and make no contribution to the event covariance.

\section{Perfomance of \texorpdfstring{\veslt}{} significance}
Here we evaluate the performance of \veslt\ significance in the \zmumu\ and dijet channels, which nominally have zero \veslt, and in the \wenu\ channel which contains genuine \veslt.  Events in the \zmumu\ channel are selected to contain two central, isolated muons with $p_T > 20$ GeV, lying within the invariant mass window $60 < M_{\mu\mu} < 120$ GeV.  Dijet events contain at least one jet with $p_T > 400$ GeV and at least two jets with $p_T > 200$ GeV.  Events in the \wenu\ channel require one central, isolated electron with $p_T > 30$ GeV.

\begin{figure}[htb]
  \centering
  \includegraphics[width=0.3\textwidth]{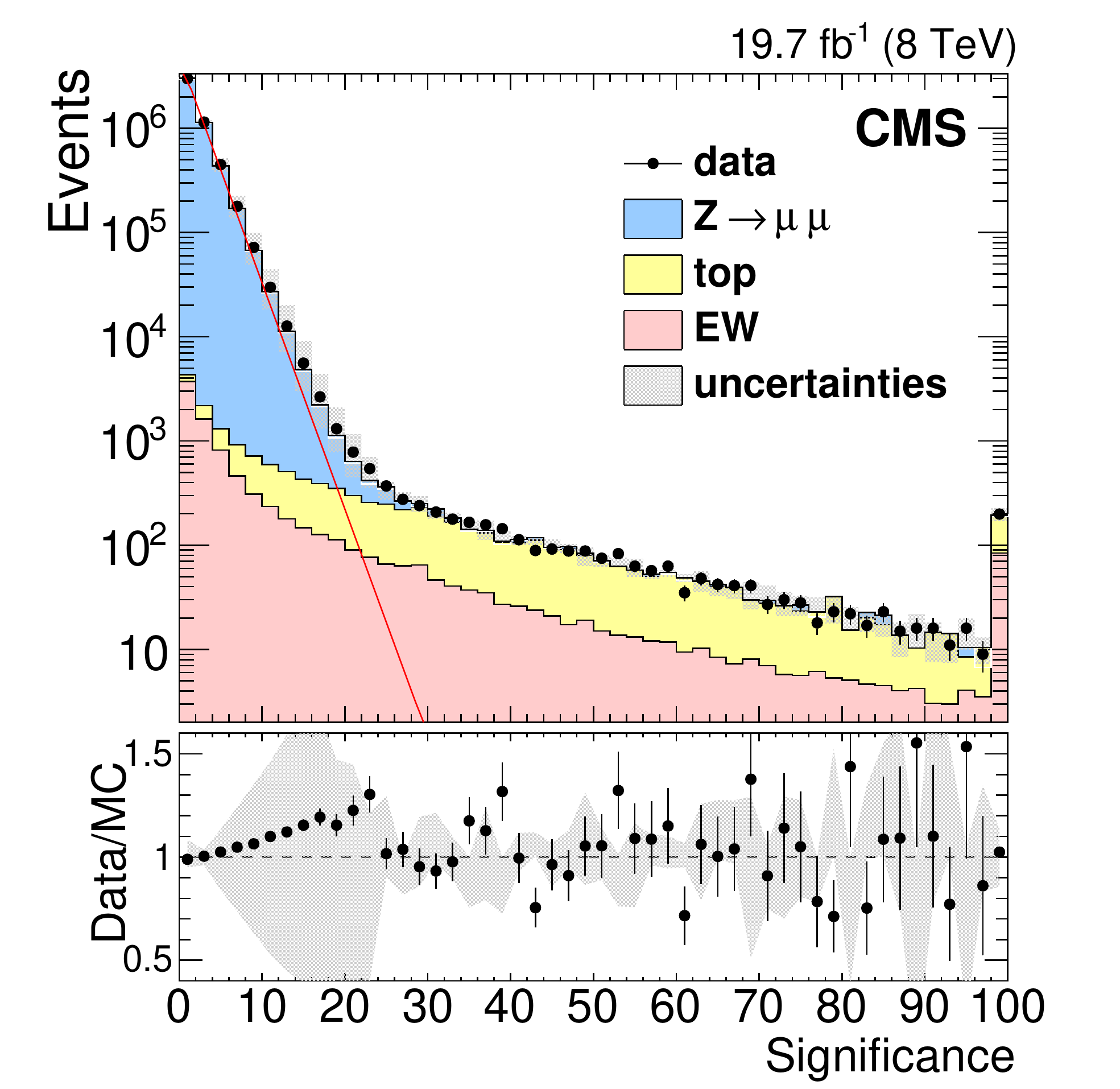}
  \includegraphics[width=0.3\textwidth]{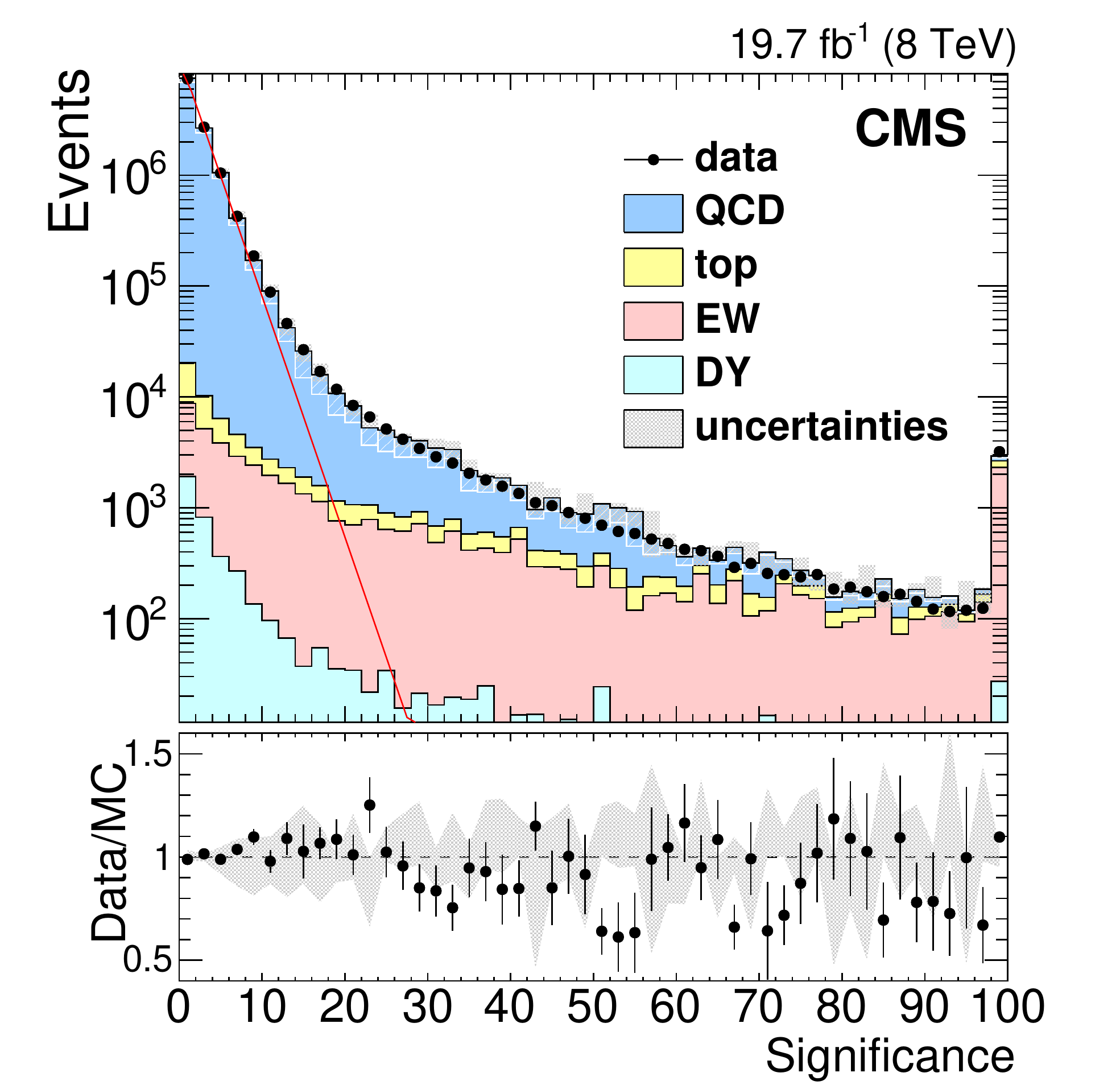}
  \includegraphics[width=0.3\textwidth]{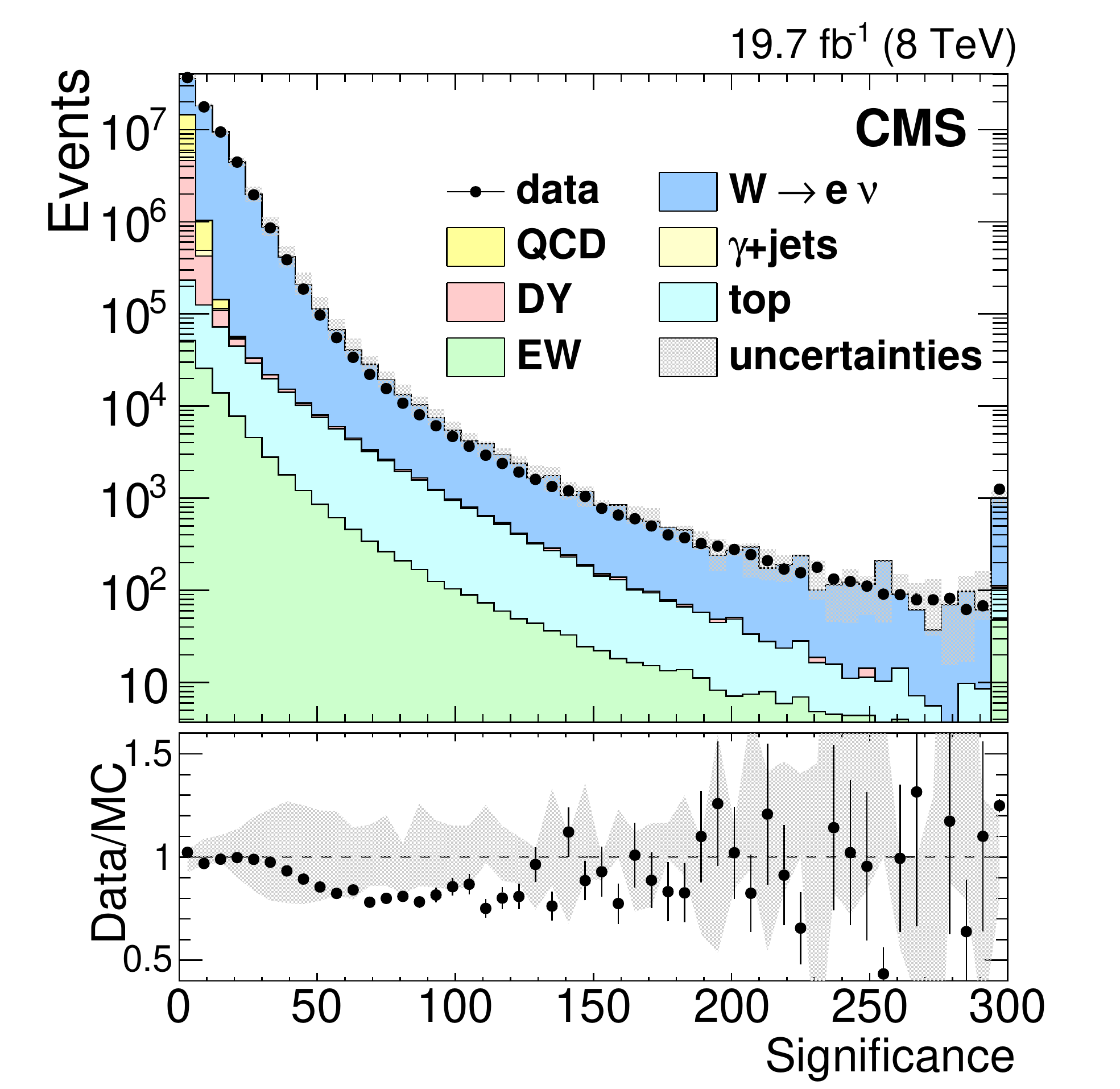}
  \caption{Distribution of \veslt\ significance in the (left) \zmumu\, (middle) dijet, and (right) \wenu\ channels.  The red line corresponds to a $\chi^2$ distribution with two degrees of freedom.  The white hashed region shows the contribution from signal events containing nonzero \veslt.}
  \label{fig:sig}
\end{figure}

\begin{figure}[htb]
  \centering
  \includegraphics[width=0.3\textwidth]{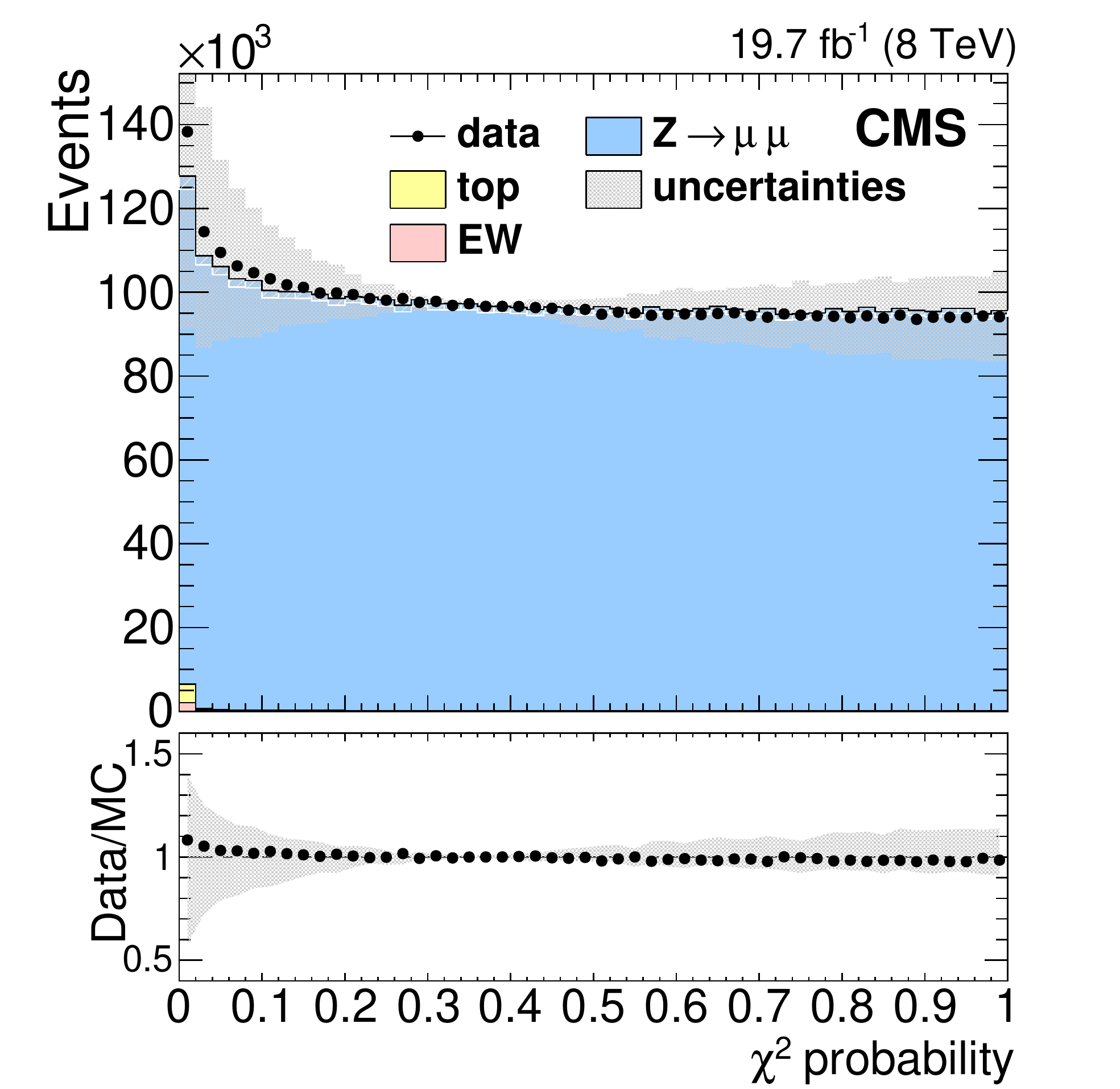}
  \includegraphics[width=0.3\textwidth]{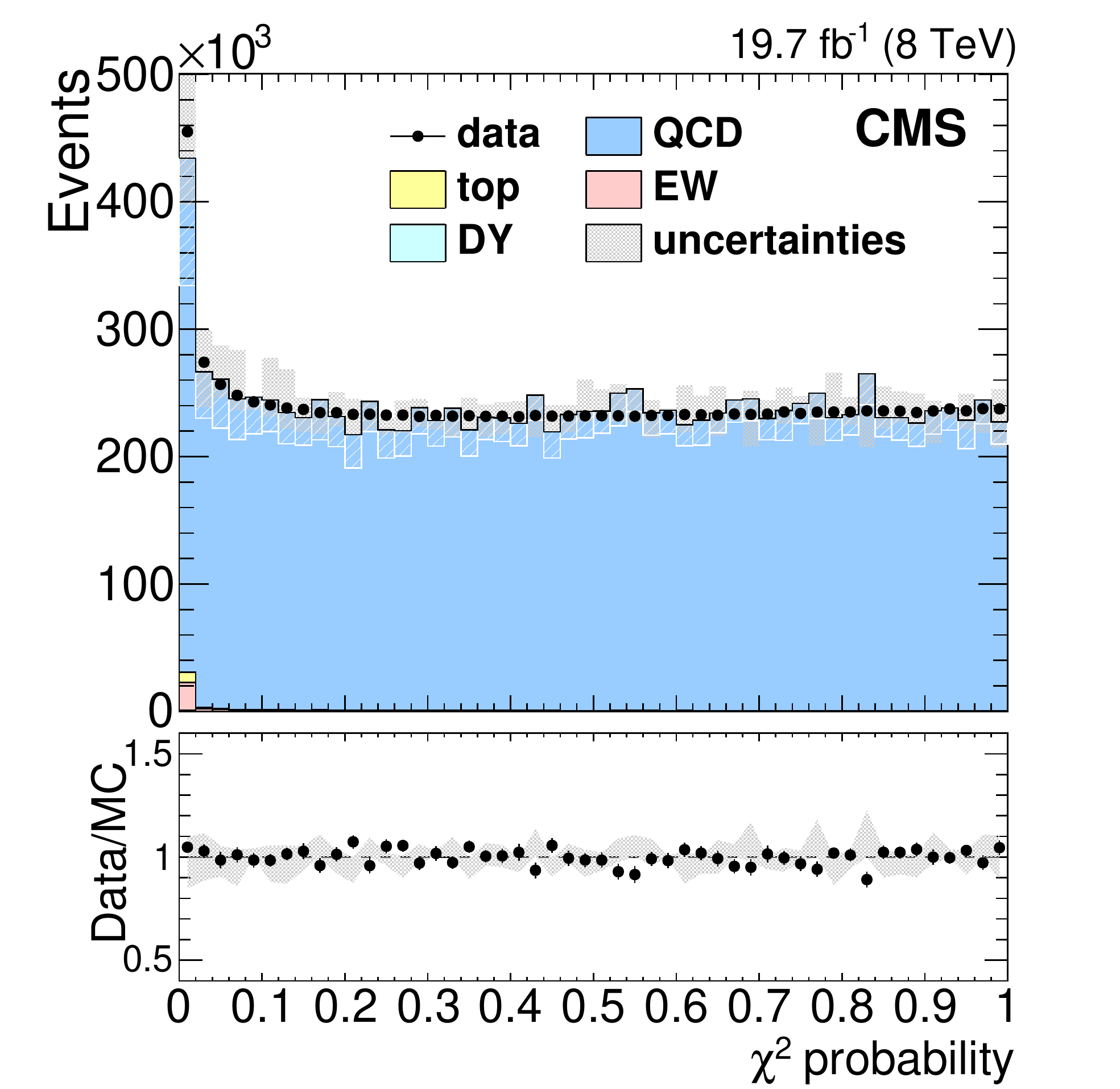}
  \includegraphics[width=0.3\textwidth]{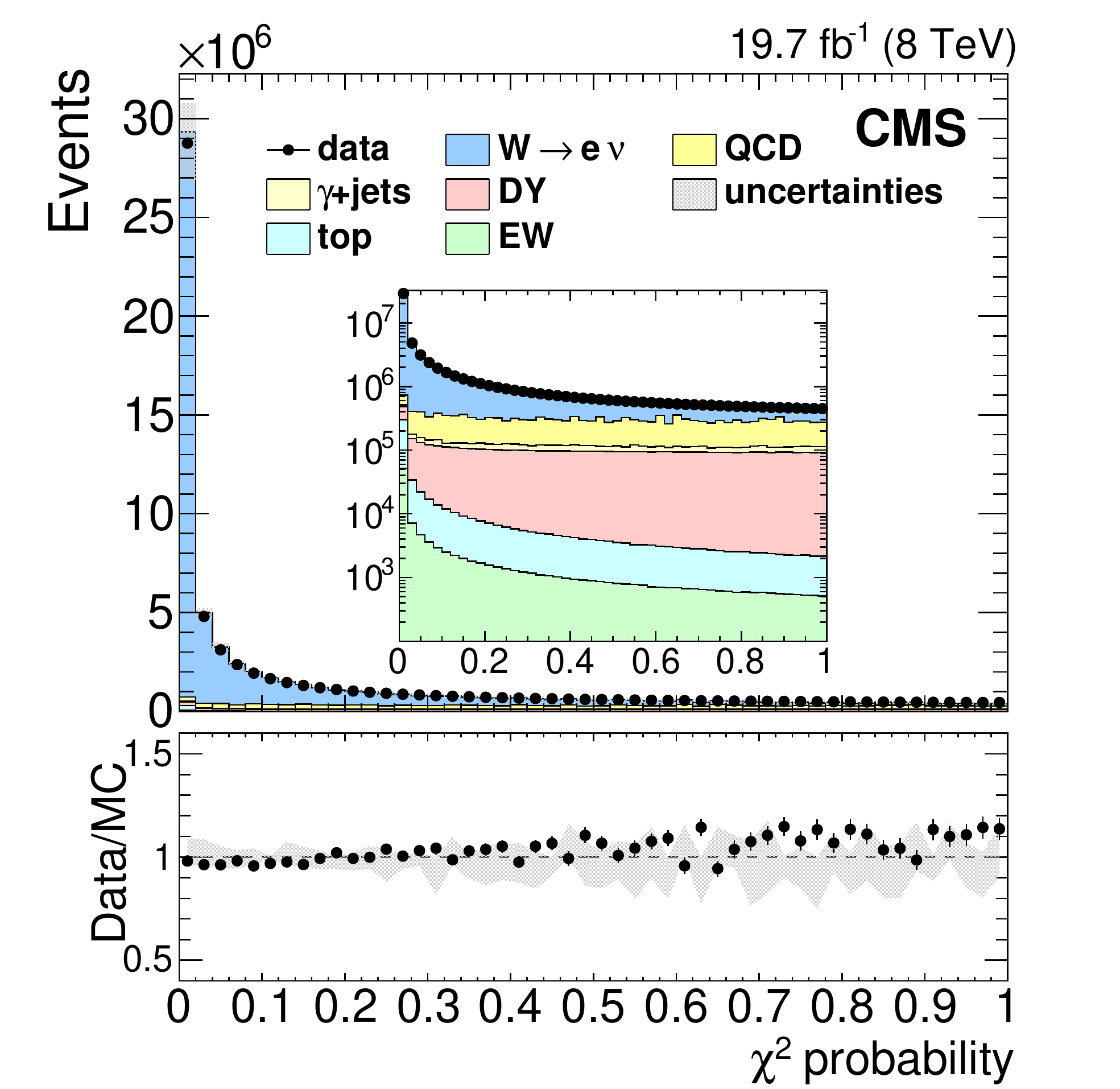}
  \caption{Distribution of $\chi^{2}$ probability in the (left) \zmumu\, (middle) dijet, and (right) \wenu\ channels.}
  \label{fig:pchi2}
\end{figure}

In Figs.~\ref{fig:sig}~and~\ref{fig:pchi2} we show distributions of \veslt\ significance, and the corresponding $\chi^2$ probability, \pchisq, for the three channels described above.
The core of the \veslt\ significance spectrum for the \zmumu\ and dijet samples follows an ideal $\chi^{2}$ distribution, but begins to slightly deviate from a perfect $\chi^{2}$ at high values of \signif.  The \pchisq\ distribution in these channels is mostly flat, with a slight excess at low values of probability due to events with nonzero \veslt\ and imperfections in the modeling of resolutions.  In the \wenu\ channel, events with genuine \veslt\ result in large values of significance and low values of \pchisq.  A separation is evident between the signal events, and the zero-\veslt\ backgrounds stemming from QCD and Z boson decays, which are found mostly at low values of significance with relatively flat distributions in \pchisq.

\begin{figure}[htb]
  \centering
  \includegraphics[width=0.309\textwidth]{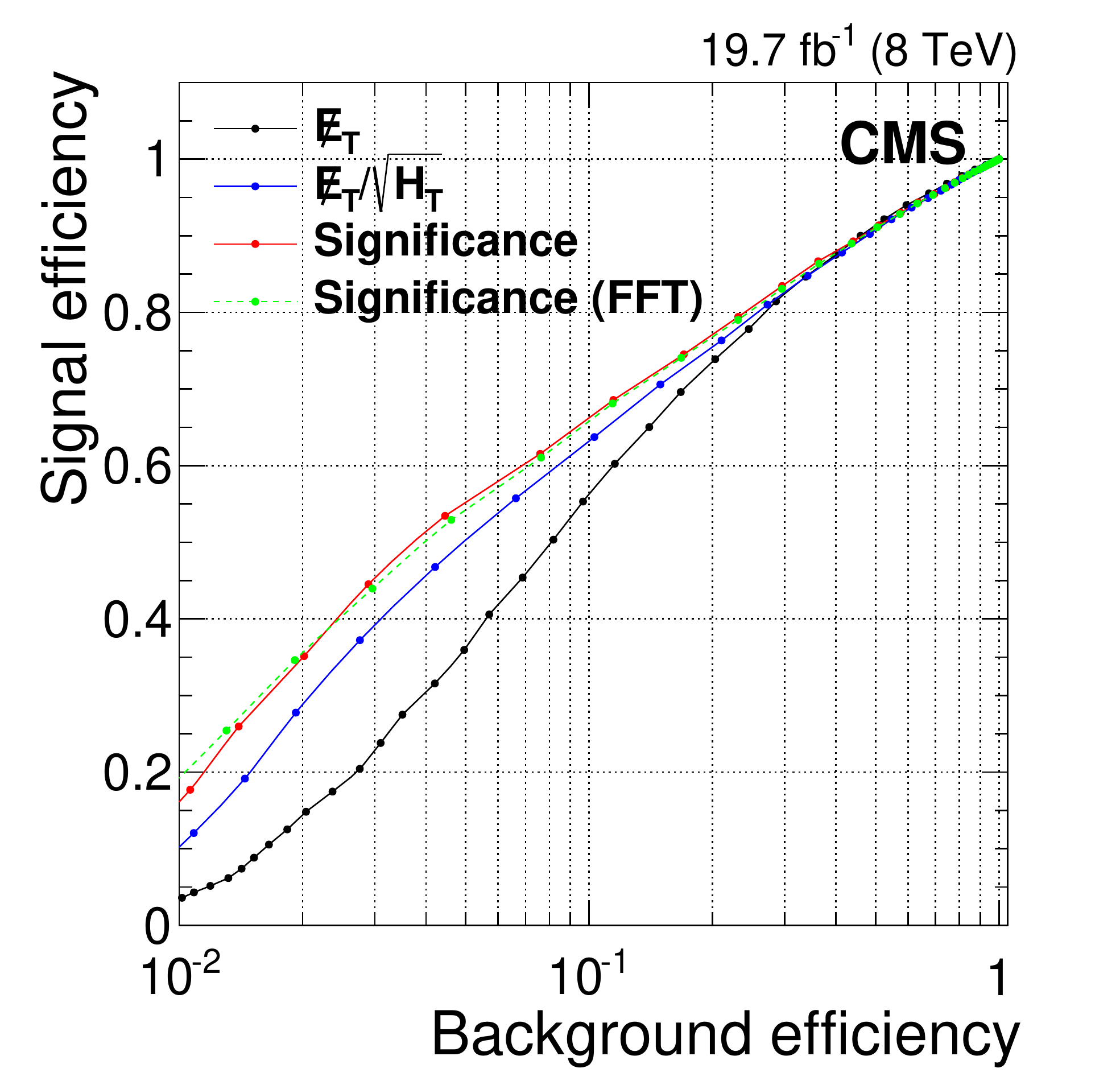}
  \includegraphics[width=0.29\textwidth]{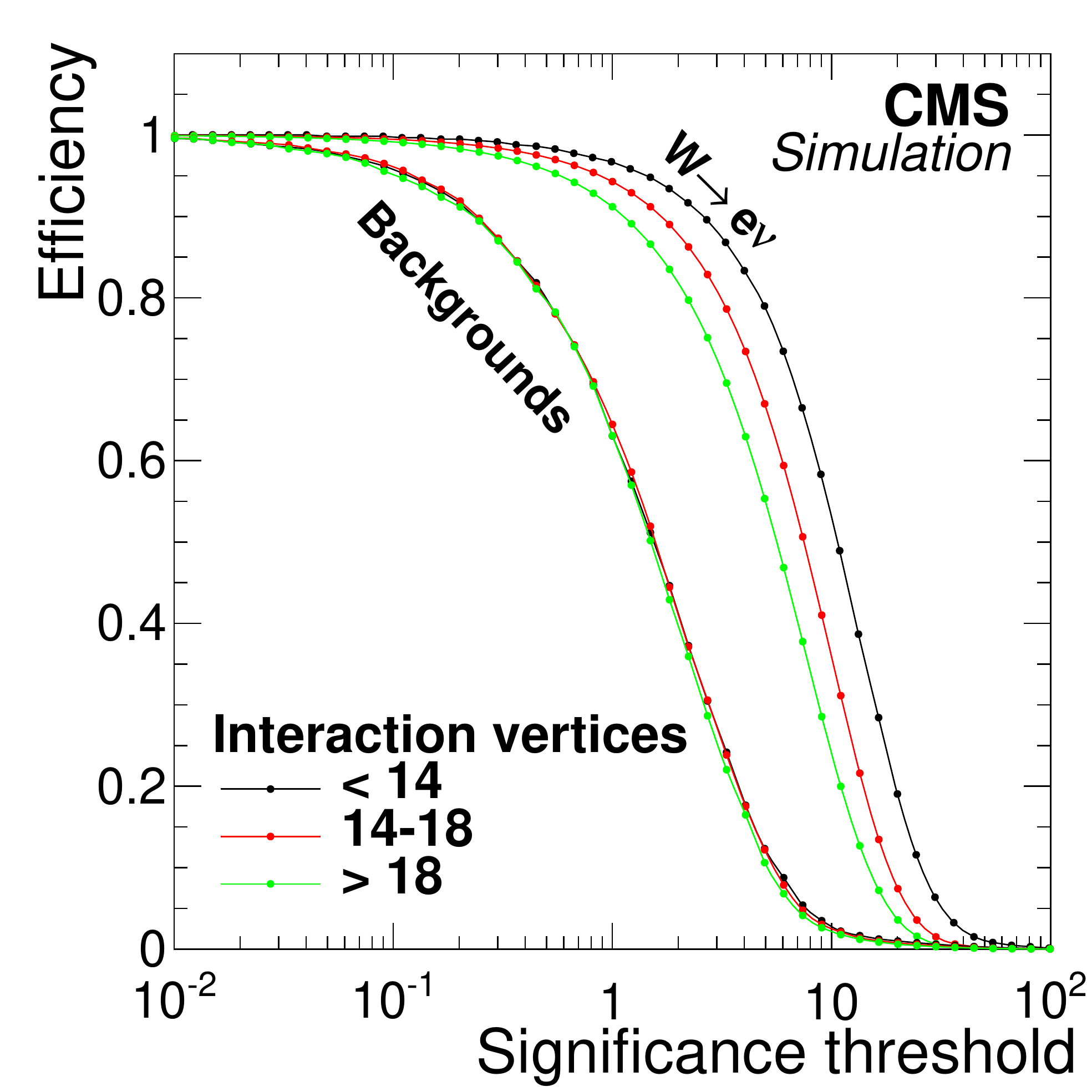}
  \includegraphics[width=0.29\textwidth]{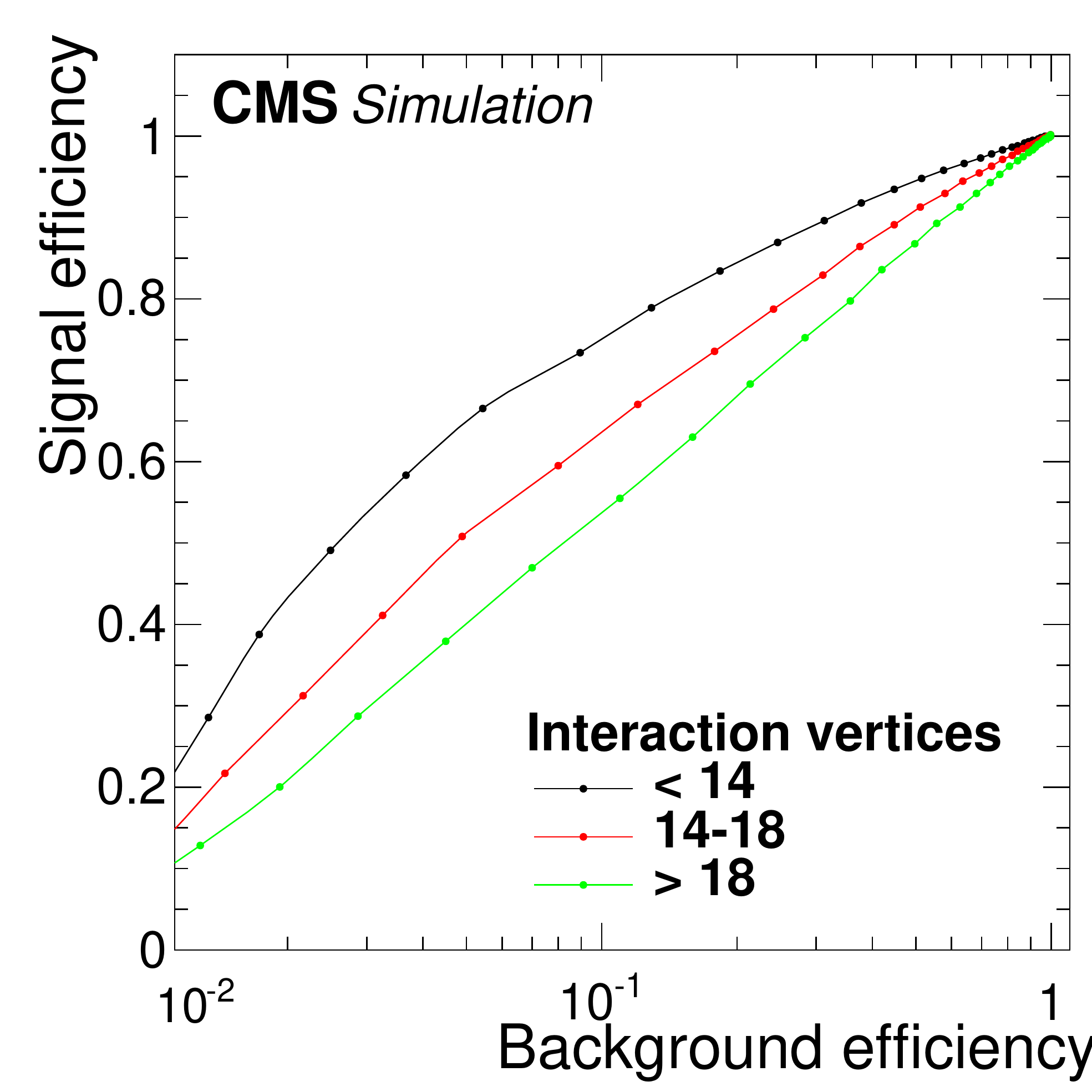}
  \caption{Signal and background efficiencies in \wenu\ events.}
  \label{fig:ROC}
\end{figure}

The potential gain of introducing \veslt\ significance into the selection criteria for \wenu\ events is evaluated in Fig.~\ref{fig:ROC}.  On the left, we compare the signal and background efficiencies in simulation while placing increasing cuts on \veslt\ significance and two simpler variables, \eslt\ and $\eslt/\sqrt{\sum{E_T}}$, the latter utilizing a scalar sum over all objects in the event.  Choosing a working point at 50\% signal efficiency, \veslt\ significance gives a 4.0\% background efficiency, while \eslt\ and $\eslt/\sqrt{\sum{E_T}}$ give 8.2\% and 5.1\% background efficiencies, respectively.  The signal and background efficiencies for \veslt\ significance are evaluated in three regimes of pile up in Fig.~\ref{fig:ROC} (middle, right).  A feature of \veslt\ significance is its stability against pile up in events with zero \veslt.  However, events with genuine \veslt\ do carry a dependence on pile up, hence the decreasing signal efficiency as the number of vertices is increased.

\begin{figure}[htb]
  \centering
  \includegraphics[width=0.3\textwidth]{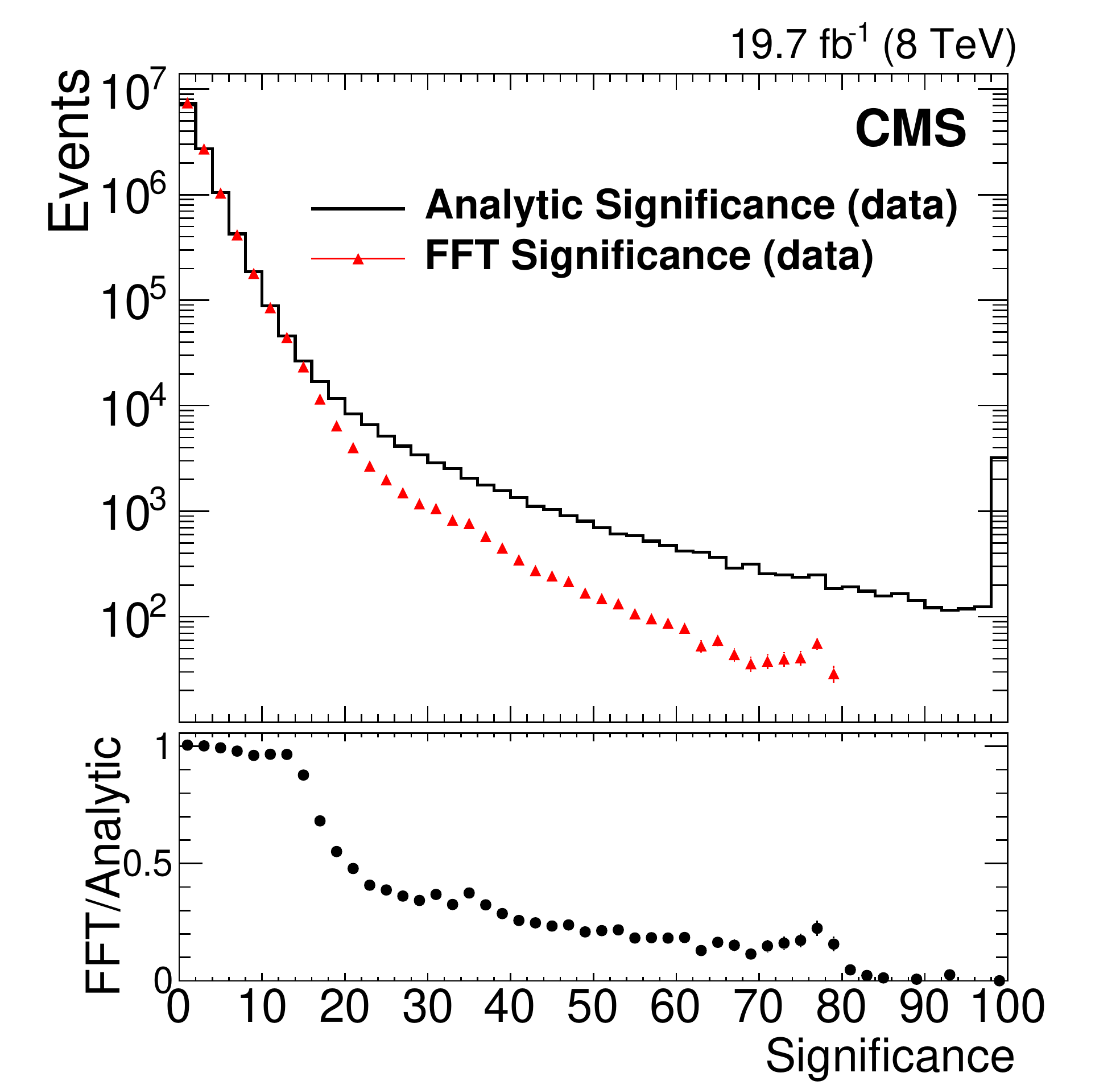}
  \includegraphics[width=0.3\textwidth]{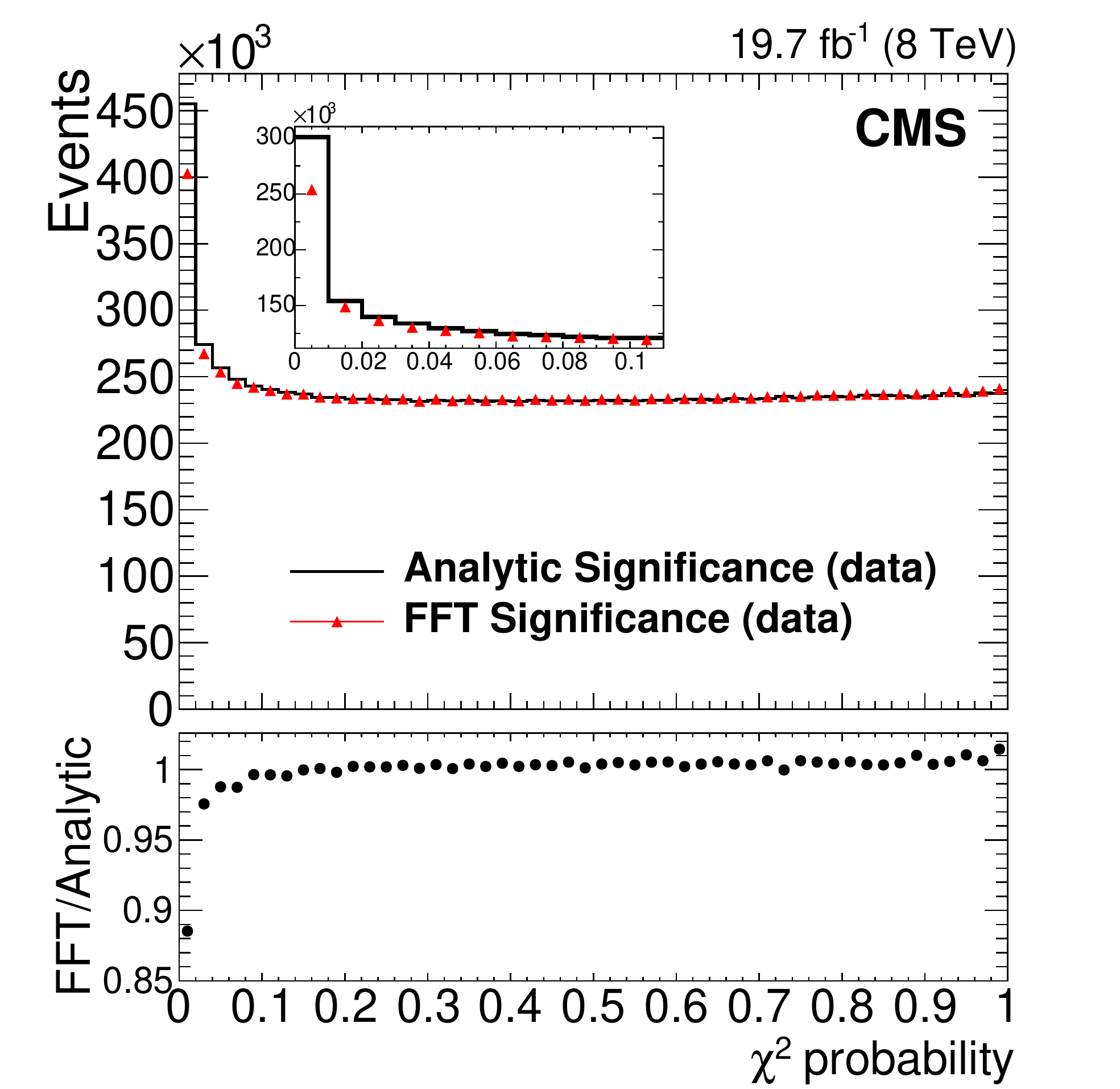}
  \caption{\veslt\ significance computed with the analytic and FFT-based approach in dijet events.}
  \label{fig:sigFFT}
\end{figure}

The \veslt\ significance variable described thus far makes the Gaussian approximation when evaluating the total \veslt\ resolution.  However, the jet $p_T$ resolution shapes are known to exhibit small non-Gaussian tails.  To include these effects, we have developed an extension to \veslt\ significance in which a Fast Fourier Transform (FFT) technique is utilized to convolve the full jet resolution shapes in each event.  This results in a \veslt\ resolution function that embodies these full jet resolution shapes, as well as the unclustered energy resolution which is assumed to be Gaussian.  The significance is then given by Eq.~\ref{eq:metsig:defn}.  In Fig.~\ref{fig:sigFFT}, the FFT-based \veslt\ significance is compared to the analytic (Gaussian) variable using dijet events in data.  The FFT significance gives a steeper fall in the tail of \signif\ and a reduced excess in the low-probability region of \pchisq, indicating a reduction of events where the resolution has been underestimated.

\section{Conclusion}
We have described the \veslt\ significance variable and assessed its performance in the full CMS 8 TeV dataset.  In the bulk of \zmumu\ and dijet events, \veslt\ significance is consistent with a $\chi^2$ variable with two degrees of freedom, and has a flat distribution in \pchisq.  In the \wenu\ channel, \veslt\ significance gives better discrimination between signal and background events when compared to simpler methods.  When computed with a FFT-based technique, \veslt\ significance can be extended include non-Gaussian effects in the jet resolution shapes.  This approach reduces the excess of high-significance events due to underestimated jet resolutions.





\end{document}